\documentclass[useAMS,usenatbib]{mnras}
\include{def_bibtex}

\usepackage{siunitx}
\usepackage{fancyhdr}
\usepackage{graphicx,latexsym}
\usepackage{color}
\usepackage{xcolor}
\usepackage{mathtools}
\usepackage[margins]{trackchanges}
\usepackage{multirow}
\usepackage{subcaption}
\usepackage{float}
\usepackage{ctable}
\usepackage[hyphenbreaks]{breakurl}
\usepackage{array}
\usepackage{color,soul}
\usepackage[section]{placeins}
\usepackage{bigints}
\usepackage{hyperref}
\usepackage{bm}
\usepackage{amsmath,amssymb}
\usepackage{pdflscape}
\usepackage{scrhack}
\usepackage{tabularx}
\usepackage{booktabs}
\usepackage{siunitx}
\usepackage{colortbl}

\newcommand{\GG}[1]{}
\addeditor{LdB}
\usepackage{times}

\newcolumntype{M}{>{\centering\arraybackslash}m{\dimexpr.35\linewidth-2\tabcolsep}}
\newcolumntype{N}{>{\centering\arraybackslash}m{\dimexpr.16\linewidth-2\tabcolsep}}
\newcolumntype{Z}{>{\centering\arraybackslash}m{\dimexpr.23\linewidth-2\tabcolsep}}
\newcolumntype{B}{>{\centering\arraybackslash}m{\dimexpr.20\linewidth-2\tabcolsep}}

\title[VLF QPOs from the 11 Hz pulsar in Terzan 5]{Very low-frequency oscillations from the 11 Hz pulsar in Terzan 5: frame-dragging back on the table.}
\author[du Buisson et al.]{L. du Buisson$^{1}$\thanks{E-mail:
lise.dubuisson@chch.ox.ac.uk}, S. Motta$^{1}$ and R. Fender$^{1,2}$\\
$^{1}$Department of Physics, University of Oxford, Denys Wilkinson Building, Keble Road, Oxford, OX1 3RH, UK.\\
$^{2}$Department of Astronomy, University of Cape Town, R. W. James Building, Rondebosch, Cape Town, 7700, South Africa.\\}

\begin{document}

\date{Accepted XXX. Received XXX; in original form XXX}

\pagerange{\pageref{firstpage}--\pageref{lastpage}} \pubyear{2018}

\maketitle

\label{firstpage}

\begin{abstract}
We present a re-analysis of 47 Rossi X-ray Timing Explorer observations of the 11Hz accreting pulsar IGR J17480-2446 in Terzan 5 during its 2010 outburst. We studied the fast-time variability properties of the source and searched for quasi-periodic oscillations (QPOs) in a large frequency range. General Relativity predicts that frame-dragging occurs in the vicinity of a spinning compact object and induces the precession of matter orbiting said object. The relativistic precession model predicts that this frame-dragging can be observed as QPOs with a characteristic frequency in the light curves of accreting compact objects. Such QPOs have historically been classified as horizontal branch oscillations in neutron star systems, and for a neutron star spinning at $11$ Hz these oscillations are expected at frequencies below $1$ Hz. However, previous studies of IGR J17480-2446 have classified QPOs at $35-50$ Hz as horizontal branch oscillations, thus casting doubts on the frame-dragging nature of such QPOs. Here we report the detection of seven very low-frequency QPOs, previously undetected, with centroid frequencies below $0.3$ Hz, and which can be ascribed to frame-dragging. We also discuss the possible nature of the QPOs detected at $35-50$ Hz in this alternative scenario. 

\end{abstract}


\section{Introduction}
\label{sec:chap4_intro}

Low-mass X-ray binaries (LMXBs) are systems containing either a neutron star (NS) or a black hole (BH) that accretes mass from a companion star. Accretion happens via an accretion disc, which produces radiation with a characteristic spectrum peaking in the X-rays \citep{Shakura1973}. 
NS LMXBs have historically been classified as either {\it Atolls} or {\it Z-sources}, based on the patterns they trace in their Colour-Colour diagrams or Hardness-Intensity diagrams (CCDs and HIDs, \citealt{Hasinger1989}), and later connected with the average accretion rates typically observed in either class: very high (often super-Eddington) for Z-sources, and relatively low (below 50\% Eddington) for Atolls (\citealt{Homan2007}). 
\cite{Munoz2014} showed that the same state/transition scheme typically used for BH systems (via the HID, \citealt{Homan2001}, and the rms-intensity diagrams, RIDs, \citealt{Munoz2011})  is evident in NS systems as well, which show hard, intermediate and soft states very similar to those in BH systems. 

The X-ray power density spectra (PDS) of both Atolls and Z-sources evolve along the HID track showing different types of narrow features superposed, usually, on broad-band noise, called quasi-periodic oscillations (QPOs). NS LMXB QPOs have been divided into high- and low-frequency QPOs. Low-frequency QPOs (LF QPOs) have centroid frequencies ranging between $\sim 0.1$ Hz and $\sim 60$ Hz. For Z-sources, LF QPOs have been historically divided into normal-branch oscillations (NBOs, \citealt{middleditch1986}), horizontal-branch oscillations (HBOs, \citealt{vanderklis1985}) and flaring-branch oscillations (FBOs, \citealt{Klis1989}) based on where they are detected along the CCD. 
Similar QPOs have been found in Atoll sources (see e.g. \citealt{DiSalvo2003}), which were divided into HBO-like and FBO-like QPOs \citep{Motta2017} in analogy with the oscillations typical of Z-sources (note that there are no NBO-like QPOs in Atoll sources). 
High-frequency QPOs (HF QPOs, \citealt{Stroh2001, belloni2012}) are called kHz QPOs in NS systems. These often appear in pairs, and thus are divided into upper and lower kHz QPOs, both with centroid frequencies spanning the range between a few hundred hertz and over a thousand hertz \citep{vanderklis1996}. 

X-ray QPOs originate in the innermost regions of the accretion flow, and are believed to be related to the geometry and dynamics of the accretion flow \citep{bookvanderklis}. However, despite being known for decades now \citep{patterson1977, Ingram2020}, QPOs remain poorly understood, and their physical origin is still largely debated.
There are different groups of suggested QPO mechanisms in the literature, all somewhat involving either the characteristic motion of matter in a strong field regime, or the oscillation of different parts of the accretion flow (see \citealt{Ingram2020} for a recent review). The relativistic precession model (RPM), originally proposed by \cite{Stella1998}, considers the motion of matter with elliptical orbits slightly tilted with respect to the spin of a compact object, and associates such motions to specific types of QPOs visible in the PDS. In particular, the nodal precession ascribed to the motion of matter related to the frame-dragging occurring around spinning compact objects - known as the Lense-Thirring (LT) effect \citep{Bardeen1975} - has been associated to the HBO and HBO-like QPOs observed in NS LMXBs. The precession of the elliptical orbit's semi-major axis - the periastron precession - and the orbital frequency are instead associated to the lower and upper kHz QPOs, respectively. 

The RPM has been used to interpret QPOs in both NS  \citep{stella1999, done1, duBuisson2020} and BH systems \citep{m2014, Motta2014b}. A remarkable exception is however represented by a NS LMXB located in the globular cluster Terzan 5, the $11$ Hz accreting pulsar IGR J17480-2446, which seems to show QPOs that cannot be explained in terms of LT precession (\citealt{terzan5}, ALT2012 from here onwards). ALT2012 analysed the fast-time variability of the source using data from the Rossi X-ray Timing Explorer (RXTE)\footnote{RXTE: \burl{https://heasarc.gsfc.nasa.gov/docs/xte/}}, and found \mbox{LF QPOs} in the range $\sim 35-50$ Hz in $6$ observations, and kHz QPOs in the range $\sim 800-920$ Hz, which in some cases appeared simultaneously with the LF QPOs. These authors classified the LF QPOs as HBOs, and showed that given the slow spin rate of the NS in the system, they could not be interpreted as the effect of LT precession, which should instead result in QPOs at frequencies strictly below $0.82$ Hz, which were however not detected. These findings thus cast doubt on the LT interpretation of HBOs in NS LMXBs and, by extension, on the RPM.

In this paper we present the results of a new analysis of the RXTE observations of \mbox{IGR J17480-2446} during its 2010 outburst. We systematically searched for QPOs at very low frequencies in the data from the Proportional Counter Array (PCA, \citealt{Jahoda2006}) in order to reassess the presence or absence of HBOs. We also searched for QPOs at up to a few thousand hertz in order to reproduce the findings by ALT2012 and allow a direct comparison to be made. 
The paper is structured as follows: Section \ref{sec:chap4_rpm} summarises the RPM while Section \ref{sec:chap4_data} details our sample selection and data analysis. In \mbox{Section \ref{sec:chap4_results}} we describe our results, in Section \ref{sec:chap4_disc} we discuss them, and Section \ref{sec:chap4_conc} summarises and concludes the study.



\section{The Relativistic precession model}
\label{sec:chap4_rpm}

The RPM was originally proposed by \cite{Stella1998}, and later revisited by the same authors as well as by others (\citealt{stella1999}, \citealt{Stella1999a}, \citealt{Merloni1999}, \citealt{Ingram2009}, \citealt{m2014}, \citealt{IngramRPM2014}). According to the RPM, the characteristic motions of a test-particle following an elliptical orbit slightly tilted out of the equatorial plane of a spinning, compact object can be associated with specific types of QPOs, which are detected in the light curves and PDS of accreting BHs and NSs.
The nodal precession of the orbits - driven by the LT mechanism \citep{Bardeen1975} - has a characteristic frequency given by $\nu_\textrm{nod} = \nu_\phi - \nu_\theta$, where $\nu_\phi$ is the orbital frequency and $\nu_\theta$ the vertical frequency. The precession of the elliptical orbit's semi-major axis (periastron precession) occurs instead at a frequency $\nu_\textrm{per} = \nu_\phi - \nu_\textrm{r}$, where $\nu_\textrm{r}$ is the radial epicyclic frequency. For the explicit forms of the above frequencies, the reader should refer to \cite{Merloni1999} and \cite{m2014}.

The RPM associates $\nu_\textrm{nod}$, $\nu_\textrm{per}$ and $\nu_\phi$ each with a respective QPO. In NS LMXBs, in particular,  $\nu_\textrm{nod}$, $\nu_\textrm{per}$ and $\nu_\phi$ are associated with the HBO, and the lower and upper kHz QPO, respectively (\citealt{Stella1998}).
\cite{m2014} showed that under the assumption that when observed simultaneously, the three QPOs relevant for the RPM are generated at the same emission radius, the three equations of the RPM form a system of three unknown parameters: the mass $M$ and spin $a$ of the compact object, and the emission radius $r$. This system can be solved analytically if all three QPOs are present, or can be used to put constraints on any of the unknown parameters if the frequency of one of the QPOs is unknown \citep{IngramRPM2014}. This approach has been followed by \cite{Franchini2017} to set constraints on the spin parameters of a sample of accreting LMXBs. 

\section{Observations and data analysis}
\label{sec:chap4_data}
We analysed all publicly available archival RXTE observations of the 11 Hz accreting pulsar IGR J17480-2446 during its 2010 outburst, consisting of 47 observations. All observations had source count rates above 10 cts/s/PCU\footnote{PCU: Proportional Counter Unit}, ensuring adequately high signal-to-noise ratios (S/N) for the subsequent analysis.

For each observation we considered \textsc{Binned}, \textsc{Event}, \textsc{Single Bit} and \textsc{Good Xenon} PCA data modes \citep{jahoda1996, Bradt1993} and calculated the PDS using a custom software under IDL\footnote{GHATS: \hspace{1mm} \burl{http://www.brera.inaf.it/utenti/belloni/GHATS_Package/Home.html}}. We used a maximum time resolution of \mbox{$1/8192$ s} ($\sim 122 \hspace{1mm} \mu s$), and divided each observation into segments of both 16 and 512 seconds, respectively, for average PDS production (see \textit{Case 1} and \textit{Case 2} below). We excluded from the analyses in \textit{Case 1} and \textit{Case 2} short observations which contained fewer than five segments, and averaged the Leahy-normalised PDS created from each segment to produce one averaged PDS per observation with a Nyquist frequency of \mbox{$4096$ Hz}, or two or more average PDS in the case of observations longer than $12000$s. We did not subtract the contribution of the Poisson noise \textit{a priori}, but fitted it when modeling the source PDS.  
We note that observation 95437-01-01-00 contains a lunar eclipse which was cut from the observation before any further analysis (\citealt{Motta2011}, \citealt{Riggio2012}). In cases of sudden drops in an observation's count rate (usually towards the beginning or end of an observation) that can be ascribed to the re-pointing of the satellite, the beginning/end of an observation is clipped away to prevent the inclusion of low-quality data. It should be noted that all data gaps are removed prior to any FFT being carried out. Our method of PDS production differed depending on the type of QPO we were attempting to detect, as described in \textit{Case 1} and \textit{Case 2} below.


We also computed the HID for the 47 observations of \mbox{IGR J17480-2446} we consider here (see Figure \ref{fig:chap4_hid}). The count rates necessary for the computation of the HID were obtained using energy spectra extracted from  \textsc{Standard 2} data, and only using PCU \mbox{unit 2}. 
For each observation, the source intensity was measured in the \mbox{$2-16$ keV} energy band, while the hardness was calculated as the ratio of counts in two energy bands as $H_{\textrm{HID}} = A/B$, where $A$ stretches between \mbox{$6-10$ keV} and $B$ stretches between $4-6$ keV (\textsc{Standard 2} channels 6-9 and 11-19, respectively). 

\subsection{Case 1: searching for very low frequency QPOs}
\label{sec:chap4_case1}

Following the prescriptions of the RPM, ALT2012 calculated that the HBO QPO of IGR J17480-2446 should fall strictly below \mbox{$0.82$ Hz}. In order to detect such very low frequency (VLF) QPOs, it is necessary to calculate PDS with a frequency resolution of \mbox{$0.08$ Hz} or better. We therefore divided our observations into segments of 512 seconds, from which we calculated PDS with a frequency resolution of \mbox{$1/512$ Hz $\sim 0.002$ Hz}, sufficient for the task at hand. 

The observations of \mbox{IGR J17480-2446} contain, however, a large number of Type-I X-ray bursts (see \citealt{Motta2011}). These bursts have a soft, thermal spectrum, but their short recurrence times (down to about 200s)  introduces quasi-periodic variability in the light curve  that can take the form of a QPO in the PDS (in this case generated from the NS surface) that is not easily distinguishable from other (accretion-driven) types of QPOs, thereby obstructing the process of finding VLF QPOs. 
The short Type-I X-ray burst recurrence time implies that - in most cases - cutting the bursts out of our observations leaves too little data for proper analysis. In other words, apart from for a few cases, it was impossible to recover data stretches long enough to reach the frequency resolution needed for our analysis.

We therefore instead performed an energy selection on our data, considering only photons in higher energy bands. In order to select bands that sufficiently removed Type-I X-ray bursts, we applied the following cuts to the entire dataset and inspected the resulting light curves:

\begin{itemize}
\item $0-120$ keV \,\,\,\, (absolute PCA channels $0-249$)
\item $8-120$ keV \,\,\,\, (absolute PCA channels $20-249$)
\item $10-120$ keV \, (absolute PCA channels $25-249$)
\item $12-120$ keV \, (absolute PCA channels $30-249$)
\item $15-120$ keV \, (absolute PCA channels $35-249$)
\item $17-120$ keV \, (absolute PCA channels $40-249$).
\end{itemize}

\noindent An example of the effect of these cuts on an observation's light curve can be seen in Appendix \ref{AppA}. From this assessment, it was found that cuts associated to the $15-120$ keV and $17-120$ keV bands adequately removed the Type-I X-ray bursts. We finally decided to use the $\sim 15-120$ keV band (absolute PCA channels 35 to 249) to derive our PDS, as this still allowed for a high enough S/N in the PDS for the subsequent analysis. We also note that the accretion-driven aperiodic and quasi-periodic variability tend to have a hard spectrum (e.g., \citealt{Sobolewska2006}), meaning our strategy effectively reduces the number of soft, non-variable photons, thus emphasising the remaining variability. Following the described strategy, we effectively minimise the contribution of Type-I X-ray bursts in our PDS while retaining possible HBO QPOs. 

The recurrence time of the Type-I X-ray bursts in our observations vary between $200$s to over $500$s, depending on the observation. Their presence could therefore generate peaks in PDS at frequencies smaller than $0.005$ Hz. We thus conservatively exclude all significant QPO-like features falling below $0.009$ Hz from our analysis (which could still be due to residual X-ray burst contributions) in order to avoid any possible contamination of our results. 

\subsection{Case 2: searching for low frequency QPOs and kHz QPOs}
\label{sec:chap4_case2}

ALT2012 found LF QPOs in the $\sim 35-50$ Hz range, and \mbox{kHz QPOs} between $\sim 800$ Hz and $920$ Hz. In order to find these features in the observations considered here, we divided each of our observations into intervals of 16 seconds for the calculation of PDS, resulting in a frequency resolution of $0.0625$ Hz. This is preferable to the 512s intervals used for \textit{Case 1}, as the large number of PDS produced in this way is averaged, significantly increasing the S/N. However, it can happen that the averaged PDS from long observations contain broadened features due to the movement of QPOs in frequency as time progresses. For the cases where an observation was longer than $12000$s, we split the observation into shorter segments of approximately $3500$s in length, and calculated an averaged PDS for each of them, which we fitted individually. 

We then investigated three different methods of PDS production to determine which one resulted in the optimal detection of \mbox{LF QPOs} and kHz QPOs. First, we simply derived the PDS using the energy band $\sim 2 - 120$ keV (absolute PCA channels 0 to 249). 
Next we used the same energy band, but also cut the Type-I X-ray bursts out of observations. The short time intervals used to calculate PDS allowed us to generate a good S/N for the average PDS using the time intervals between consecutive Type-I X-ray bursts.  
Finally, we produced PDS in the  energy band \mbox{$\sim 15-120$ keV} (absolute PCA channels 35 to 249) without any burst cuts. 
The first method presented us with the largest number of significant QPOs in our data, and it was therefore employed to search for LF QPOs and kHz QPOs.

\subsection{Power spectral fitting}
\label{subsec_chap4:selection}

To find the QPOs present in our dataset, we preselected the PDS of observations that visually contained these features for each of the cases above. The features of each power spectrum were fit with a combination of Lorentzians and a power-law component (to account for the Poisson noise) by means of the XSPEC package (\citealt{Arnaud1996}) by using a one-to-one energy-frequency conversion for our PDS and a unity response matrix. We excluded all non-significant features from the analysis: for very low frequencies (where flat-top noise or red noise were present), this meant excluding features below a significance\footnote{Significance is calculated as the integral of the power of the Lorentzian used for the fitting of the feature divided by the negative $1\sigma$ error on this integral.} of $2\sigma$ (in order to account for the lower relative frequency resolution); at all other frequencies features had to be significant at or above $3\sigma$. QPOs were identified by requiring that a given feature is detected at a significance of $3\sigma$ or above, and has a quality factor $Q \geq 2$ (taking uncertainties into account). Here $Q = v_c / \Delta \nu$, where $v_c$ is the centre frequency and $\Delta \nu$ the FWHM\footnote{FWHM: full width at half maximum} of the Lorentzian.

Due to the X-ray pulsar nature of the NS in \mbox{IGR J17480-2446}, a very narrow peak corresponding to the  11 Hz pulsation is visible in our PDS. We do not cut this peak from the PDS - instead, we fit it along with the rest of the features. 





\section{Results}
\label{sec:chap4_results}

\begin{figure}
\centering
\includegraphics[width=0.5\textwidth]{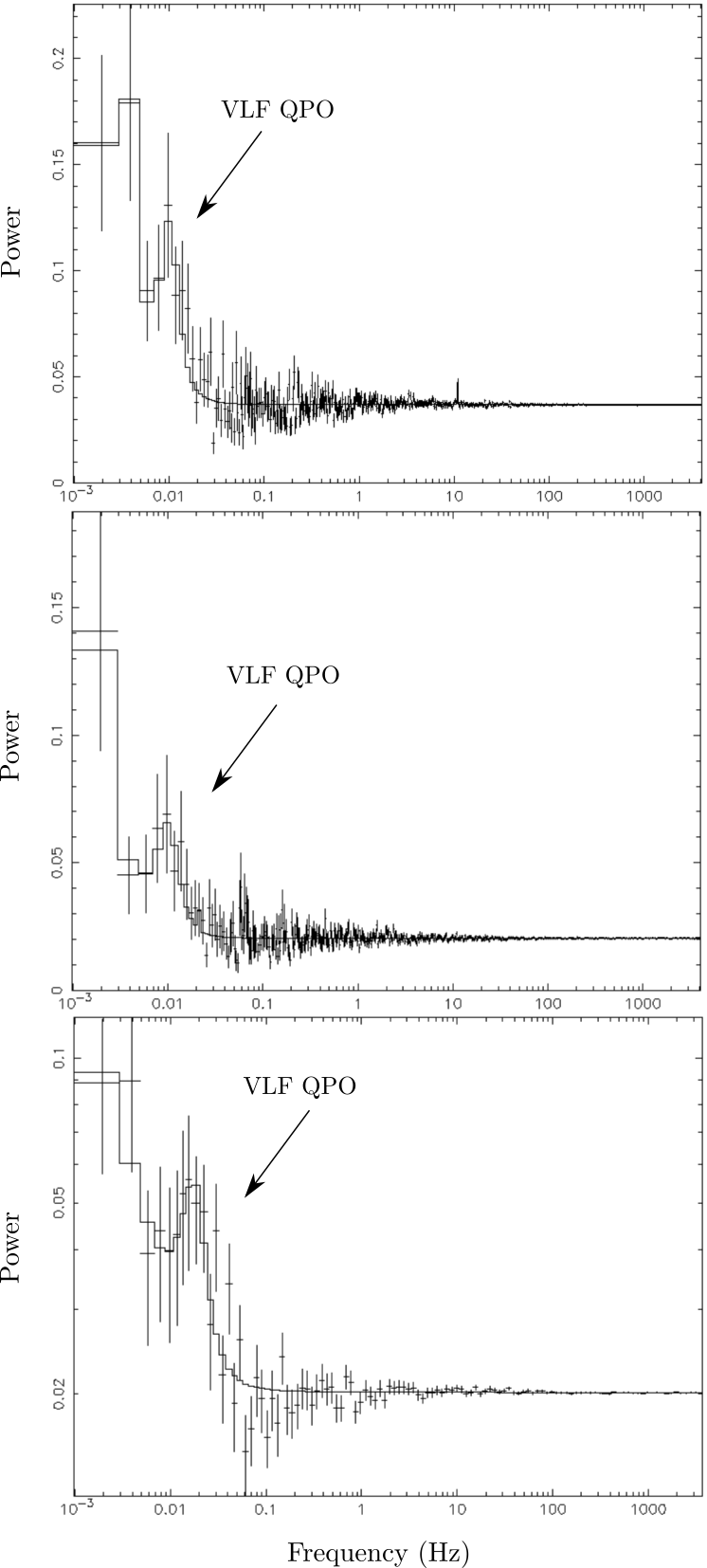}
\caption{The PDS of three observations in which we found VLF QPOs (see also Table \ref{tab:chap4_qpos}). Top: Obs ID 95437-01-02-01. Middle: Obs ID \mbox{95437-01-10-05.} Bottom: Obs ID 95437-01-11-03.}
\label{fig:vlfqpos}
\end{figure}

We report in Table \ref{tab:chap4_qpos} the VLF QPOs found by carrying out the analysis described in \mbox{Case 1} and fitting the resulting PDS (top section), as well as the LF and \mbox{kHz QPOs} found through the analysis described in Case 2 (bottom two sections). For more information on the fitting of the PDS of the observations containing VLF QPOs, see Appendix \ref{AppB}.
We found $7$ VLF QPOs having centre frequencies in the range $\sim 0.01 - 0.03$ Hz (for illustration, three of these are shown in Figure \ref{fig:vlfqpos}), $3$ \mbox{LF QPOs} in the range $\sim 44 - 50$ Hz and $3$ kHz QPOs in the range $\sim 840 - 870$ Hz. The frequency ranges of our LF and kHz QPOs fall within those reported by ALT2012. We note, however, that the analysis performed by these authors differed slightly from ours, and a list of their six observations considered is not given. It is therefore not trivial to determine if and when our detections correspond exactly to those reported in ALT2012. We further note that for the observations in which VLF QPOs were found, the recurrence times for Type-I X-ray bursts (before they were removed by our energy cuts) were $>670$s - this would translate to a peak (if any) in the PDS at $<0.0015$ Hz. Our conservative exclusion of all significant QPO-like features falling below $0.009$ Hz from our analysis (see \mbox{Section \ref{sec:chap4_case1}}) is therefore justified.


\begin{table}
\renewcommand{\arraystretch}{1.3}
\caption[QPOs]{A list of QPOs detected in our analysis. VLF QPOs were found following \textit{Case 1} analysis (see Section \ref{sec:chap4_case1}). VLF QPO observations with a star next to their Obs IDs indicate observations for which second harmonics of the VLF QPOs were also found. These are not listed in the table as they do not form part of our analysis, but more detail regarding fit parameters for observations containing VLF QPOs can be found in Appendix \ref{AppB}. LF and kHz QPOs were found through \textit{Case 2} analysis (see Section \ref{sec:chap4_case2}). In the cases where observations were cut into two or more independent segments, an extra digit attached at the end of their observation IDs indicate which segment was used. See the text for more details. All errors reported in this table are $1 \, \sigma$ errors.}
    \begin{tabular}{lccc}
    \toprule
    \textbf{Obs ID} & \textbf{Frequency (Hz)} & \textbf{Q factor} & \textbf{Significance} \\ \hline \hline
    \multicolumn{4}{c}{\textbf{VLF QPOs}} \\
    \hline \hline
    95437-01-02-01 & $0.0101^{+0.0011}_{-0.0008}$ & $1.9 \pm 0.7$ & $4.3 \, \sigma$  \\
    95437-01-10-05 & $0.010^{+0.002}_{-0.001}$ & $1.6 \pm 0.8$ & $6.2 \, \sigma$\\
    95437-01-11-03  & $0.016 \pm 0.002$ & $2 \pm 1$  & $3.2 \, \sigma$  \\
    95437-01-11-06* & $0.020 \pm 0.001$ & $4 \pm 1$ & $3.9 \, \sigma$ \\
    95437-01-12-04* & $0.0175 \pm 0.0006$ & $5 \pm 2$ & $4.2 \, \sigma$ \\
    95437-01-13-04 & $0.029 \pm 0.002$ & $3 \pm 1$ & $3.8 \, \sigma$ \\
    95437-01-14-00 & $0.014 \pm 0.002$ & $2 \pm 1$ & $3.2 \, \sigma$ \\
    
    \hline \hline
    \multicolumn{4}{c}{\textbf{LF QPOs}} \\
    \hline \hline
    95437-01-07-00 & $47.5 \pm 0.5$ & $3.2 \pm 0.4$ & $11.9 \, \sigma$ \\
    95437-01-08-00-1 & $49.2^{+0.8}_{-0.9}$ & $5 \pm 2$ & $4.8 \, \sigma$\\
    95437-01-09-00 & $44.7 \pm 0.6$ & $5 \pm 1$ & $6.9 \, \sigma$ \\
    
    \hline \hline
    \multicolumn{4}{c}{\textbf{kHz QPOs}} \\
    \hline \hline
    95437-01-07-00 & $840^{+13}_{-11}$ & $12 \pm 5$ & $3.3 \, \sigma$ \\
    95437-01-09-00 & $854 \pm 4$ & $22 \pm 8$ & $4.9 \, \sigma$ \\
    95437-01-10-01 & $870^{+6}_{-7}$ & $21 \pm 10$ & $3.7 \, \sigma$ \\
    \toprule
    \end{tabular}
    \label{tab:chap4_qpos}
\end{table}

\begin{figure}
\centering
\includegraphics[width=0.5\textwidth]{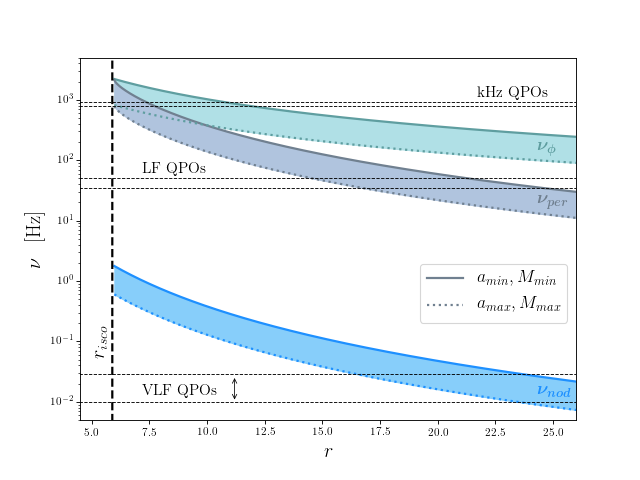}
\caption{The minimum and maximum theoretical values of $\nu_{\textrm{nod}}$, $\nu_{\textrm{per}}$ and $\nu_{\phi}$ as a function of the emission radius $r$ for IGR J17480-2446, inferred using the RPM. Coloured lines indicate the maximum and minimum predicted values, respectively. Straight horizontal dashed lines mark the frequency ranges in which QPOs were found.}
\label{fig:chap4_boundaries}
\end{figure}

In order to compare the quasi-periodic features we detected with the predictions of the RPM, we used the RPM equations given in \citealt{m2014} (see also Section \ref{sec:chap4_rpm}) to plot the theoretical estimates of $\nu_{\textrm{nod}}$, $\nu_{\textrm{per}}$ and $\nu_{\phi}$ as a function of the emission radius $r$. 
We used the measured spin frequency $\nu = 11$ Hz of the source, and calculated the minimum and maximum dimensionless spins ($a_{min}$ and $a_{max}$) by assuming a minimum and maximum moment of inertia for the NS ($I_{min}$ and $I_{max}$), respectively. To do so, we followed \cite{mukherjee2018}, and in particular their Figure 4, which shows the moment of inertia as a function of different NS masses inferred for a number of realistic candidate NS equations of state. We found $I_{min} = 0.75 \times 10^{45}$ $\textrm{g} \hspace{0.5mm} \textrm{cm}^2$ for a NS mass of $M_{min} = 1.0$ $\textrm{M}_{\odot}$, and $I_{max} = 5 \times 10^{45}$ $\textrm{g} \hspace{0.5mm} \textrm{cm}^2$ for a mass of $M_{max} = 2.7$ $\textrm{M}_{\odot}$. We then calculated the minimum and maximum dimensionless spin parameters $a_{min}$ and $a_{max}$, to find $0.0054 \leq a \leq 0.0059$. Next, we used Equations 9, 6 and 1 in \cite{IngramRPM2014} to determine the minimum and maximum theoretical estimates of $\nu_{\textrm{nod}}$, $\nu_{\textrm{per}}$ and $\nu_{\phi}$ as a function of the emission radius $r$, assuming first $a_{min}$ and $M_{min}$, and then $a_{max}$ and $M_{max}$. Our result is displayed in Figure \ref{fig:chap4_boundaries}, where we also marked the frequency ranges where we detected QPOs in our data.  

\begin{figure}
\centering
\includegraphics[width=0.5\textwidth]{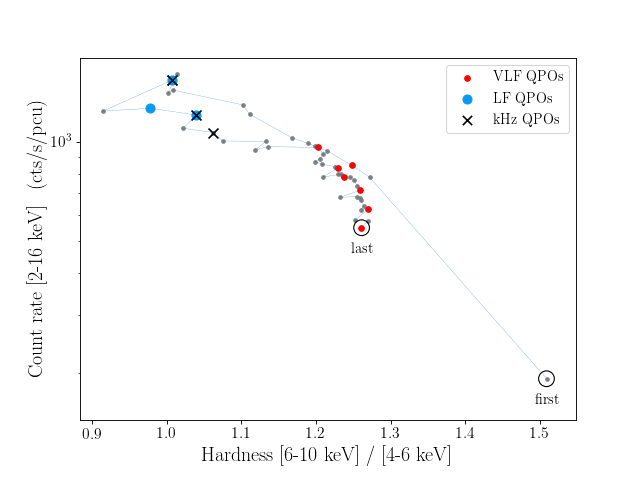}
\caption{The HID of IGR J17480-2446, where the source count rate is plotted against the spectral hardness. Each data point represents an observation, and consecutive observations are connected. The first and last observation considered are marked on the plot and denoted by `first' and `last', respectively. Small red dots are observations containing VLF QPOs, larger light blue dots are observations containing LF QPOs, and black crosses are observations containing kHz QPOs (see also \mbox{Table \ref{tab:chap4_qpos}}). Light blue dots overlaid by black crosses contain both LF QPOs and kHz QPOs. A colour version of this figure is available online.}
\label{fig:chap4_hid}
\end{figure}

The HID for our data is shown in Figure \ref{fig:chap4_hid}, showing the source count rate and hardness of each observation in our analysis. Consecutive observations are connected via thin lines, with the earliest and latest observations circled and numbered. Red data points indicate observations containing VLF QPOs, light blue observations contain LF QPOs and black observations contain kHz QPOs (note that in two cases LF QPOs and kHz QPOs are detected simultaneously). VLF QPOs are detected both at the beginning and at the end of the outburst, in a hard state, while LF QPOs and kHz QPOs are found in a relatively soft state, close to the peak of the outburst which, according to \cite{Motta2011}, reached and possibly exceeded the Eddington limit.

\section{Discussion}
\label{sec:chap4_disc}

Based on the assumption that all types of QPOs, both in NS and in BH systems, should show at least a mild dependence on the spin of the compact object, we hypothesised that given the very low spin of IGR J17480-2446 ($11$ Hz, i.e. over an order of magnitude smaller than the average spin of accreting NSs, see \citealt{vanD2017}), HBOs should be detected at very low frequencies in this system (i.e. significantly below 1 Hz), a hypothesis made by ALT2012 themselves. We therefore repeated the analysis of the RXTE observations of IGR J17480-2446, following a method specifically aimed at finding PDS features with very low centroid frequencies. We also searched for QPOs up to a few thousand hertz in order to reproduce the results of ALT2012 and to allow for a direct comparison to be made. We found $7$ VLF QPOs with centre frequencies in the range $\sim 0.01-0.03$ Hz (see Table \ref{tab:chap4_qpos}), consistent with the values predicted by ALT2012 for HBOs in a slow-spinning NS. We also found $3$ LF QPOs and $3$ kHz QPOs (see Table \ref{tab:chap4_qpos}), with centroid frequencies consistent with those reported by the same authors. 

We estimated the theoretical values of $\nu_{\textrm{nod}}$, $\nu_{\textrm{per}}$ and $\nu_{\phi}$ as a function of the radius $r$, obtained assuming a NS with mass $M$ between 1 and 2.7 $M_{\odot}$, and  adopting a dimensionless spin parameter $0.0054 \leq a \leq 0.0059$. 
By comparing our estimates with the QPOs we detected we observe the following: 

\begin{enumerate}

\item The VLF QPOs we detected are consistent with being HBOs - QPOs generated through LT precession of the accretion flow - at a radius larger than approximately $15 \hspace{1mm} R_g$ ($\approx 35$ km) from the NS centre. This is supported by the fact that these features all appear in a relatively hard state, as is clear from the HID, where the inner disc radius is believed to be truncated far from the NS surface (see e.g. \citealt{Done2007}).

\item The kHz QPOs that we detected are consistent with being either the upper or lower kHz QPOs, thus associated to the orbital or periastron precession frequency of matter at a radius lower than approximately $12 \hspace{1mm} R_g$ ($\approx 25$ km) from the NS centre. The lack of a simultaneous detection of two kHz QPOs prevents any further classification. 

\item As already noted by ALT2012, the LF QPOs detected are not consistent with the nodal precession frequency around a NS spinning at $11$ Hz. They are, in principle, consistent with the periastron precession frequency of matter orbiting at a distance of $15-25 \hspace{0.5mm} R_g$ from the NS. However, we note that two of our \mbox{LF QPOs} are detected together with kHz QPOs at $\sim 840$ Hz (see \mbox{Table \ref{tab:chap4_qpos}}, observation 95437-01-07-00 and 95437-01-09-00). According to Figure \ref{fig:chap4_boundaries}, this would imply the simultaneous detection of QPOs at two very different radii, which would constitute a violation of one of the key assumptions of the RPM.  
\end{enumerate}

Concerning point (iii) we also tested for the non-simultaneity of the two QPOs in observations 95437-01-07-00 and \mbox{95437-01-09-00} by splitting each into several smaller segments, but both features seem to be present during a large fraction of these two observations. Assuming the correctness of the RPM, the simultaneity of the kHz QPOs and LF QPOs point to two main scenarios: either that the QPO associated to the periastron precession frequency (that at $\sim 47.5$ Hz) is generated at a radius larger than that at which the orbital frequency QPO (that at $\sim 840$ Hz) is generated, thus effectively invalidating a key assumption of the RPM; or that the \mbox{LF QPOs} we detected are not associated to any of the three motions relevant for the RPM, having already stated in (iii) that they cannot be explained by LT precession. 

The first scenario pushes us toward the same conclusion drawn by ALT2012, i.e. that the RPM cannot explain the frequency properties of at least some of the QPOs, at least in its simplest form. 
The second scenario, instead, further stresses the question of the real nature of the LF QPOs observed between $35$ and $50$ Hz in \mbox{IGR J17480-2446.} ALT2012 classified the QPOs at $35-50$ Hz based on their frequency (comparable with the values typically seen in other NS systems), and based on the position these features occupied in the so-called Wijnands -- van der Klis correlation (\citealt{Wijnands1999a}), formed by the frequency of a broad PDS component (called $L_b$) and the frequency of the HBO. The small number of QPOs detected by both ALT2012 and by ourselves does not really allow for the establishing of a correlation, and the fact that most of the components identifiable in a NS PDS correlate with one another (see \citealt{Psaltis1999}) suggests that ALT2012's classification of the LF QPOs at $35-50$ Hz might not be correct. Unfortunately, given the very low frequency of the VLF QPOs that we report in this work, it is not possible to determine whether these do fit into the Wijnands \& van der Klis correlation (the $L_b$ component's centroid frequency would be visible below $10$ mHz, which is lower than our frequency resolution, assuming HBO frequencies lower than $0.1$ Hz). Only a new outburst form this source and new data will therefore allow us to reach more conclusive results.

Is there an alternative explanation for these LF QPOs? PDS from NS systems are notoriously more structured and feature-rich than those from BH systems. Among the still-largely unknown features typical of NS systems are hectohertz QPOs (see e.g. \citealt{Altamirano2008}). These QPOs have been detected in a number of Atoll sources - 4U0614+09, 4U1608-52, 4U1728-34 and \mbox{4U1636-53} - around $\sim 100$ Hz. Interestingly, all these sources contain NSs spinning at fairly high frequencies ($415$, $620$, $363$ and $581$ Hz, respectively). While there is no known correlation between the NS spin and the frequency of hectohertz QPOs, it seems plausible that in a slowly spinning NS such as the one in \mbox{IGR J17480-2446} hectohertz QPOs can appear at frequencies lower than $\sim 100$ Hz. We therefore speculate that the LF QPOs detected in IGR J17480-2446 might be relatively low-frequency hectohertz QPOs.

\section{Summary and conclusions}
\label{sec:chap4_conc}
We have examined all $47$ RXTE PCA observations of the $11$ Hz accreting pulsar IGR J17480-2446 located in the globular cluster Terzan 5 during its 2010 outburst. We searched for QPO features located between $0.01$ Hz and $\sim 4000$ Hz. 

We found $7$ VLF QPOs with centre frequencies in the range $\sim 0.01-0.03$ Hz, $3$ LF QPOs in the range $\sim 44-50$ Hz, and $3$ kHz QPOs in the range $\sim 840-870$ Hz. We compared the theoretical values of the nodal frequency $\nu_{\textrm{nod}}$, periastron precession frequency $\nu_{\textrm{per}}$ and orbital frequency $\nu_{\phi}$ as a function of the emission radius as predicted by the RPM to our findings. We find that the centroid frequencies of our detected VLF QPOs are consistent with the predicted nodal frequencies if generated at radii larger than $15 \hspace{1mm} R_g$. We also find that our LF QPOs detected at $40-50$ Hz could be consistent with the periastron precession frequency of material orbiting at approximately $15 \hspace{1mm} R_g$ from the NS centre. The presence of kHz QPOs simultaneous to these LF QPOs, however, either disproves this hypothesis or invalidates one of the main assumptions of the RPM.  

We have shown that VLF QPOs at frequencies consistent with those expected for LT driven modulations are present in the data, even though more data are required to confirm the classification of such features. While not conclusive, our results cast (even more) doubt on the nature of the $35-50$ Hz QPOs detected in \mbox{IGR J17480-2446}. These LF QPOs could either be associated with the periastron precession frequency (though in this case a key assumption of the RPM needs to be relaxed significantly), or with a type of QPO known as hectohertz QPOs, observed here at smaller frequencies possibly due to the low NS spin. It is also, of course, entirely possible that these LF QPOs are simply a new type of QPO, possibly peculiar to this unique system, or perhaps typical of slowly spinning accreting NS LMXBs, of which IGR J17480-2446 is currently the only example.


\section*{Acknowledgments}
\label{sec:chap4_acknowledgments}
LdB acknowledges support from the Rhodes Trust and Christ Church College. SEM \mbox{acknowledges} the Science and Technology Facilities Council (STFC) for financial support, and the Oxford Centre for Astrophysical Surveys, which is funded through generous support from the Hintze Family Charitable Foundation. The authors thank Diego Altamirano for valuable discussions.

\section*{Data availability}
\label{sec:data_avail}
The data underlying this article are publicly available from the RXTE Archive: \url{https://heasarc.gsfc.nasa.gov/docs/xte/archive.html}.

\bibliographystyle{mnras}
\bibliography{ms}

\appendix

\section{Case 1 energy band cuts}
\label{AppA}
In order to correctly select higher energy bands that would sufficiently remove Type-I X-ray bursts from observations in our \mbox{\textit{Case 1}} analysis, the following cuts were applied to the entire dataset:

\begin{itemize}
\item $0-120$ keV \,\,\,\, (absolute PCA channels $0-249$)
\item $8-120$ keV \,\,\,\, (absolute PCA channels $20-249$)
\item $10-120$ keV \, (absolute PCA channels $25-249$)
\item $12-120$ keV \, (absolute PCA channels $30-249$)
\item $15-120$ keV \, (absolute PCA channels $35-249$)
\item $17-120$ keV \, (absolute PCA channels $40-249$).
\end{itemize}

\begin{figure*}
\includegraphics[width=0.77\textwidth]{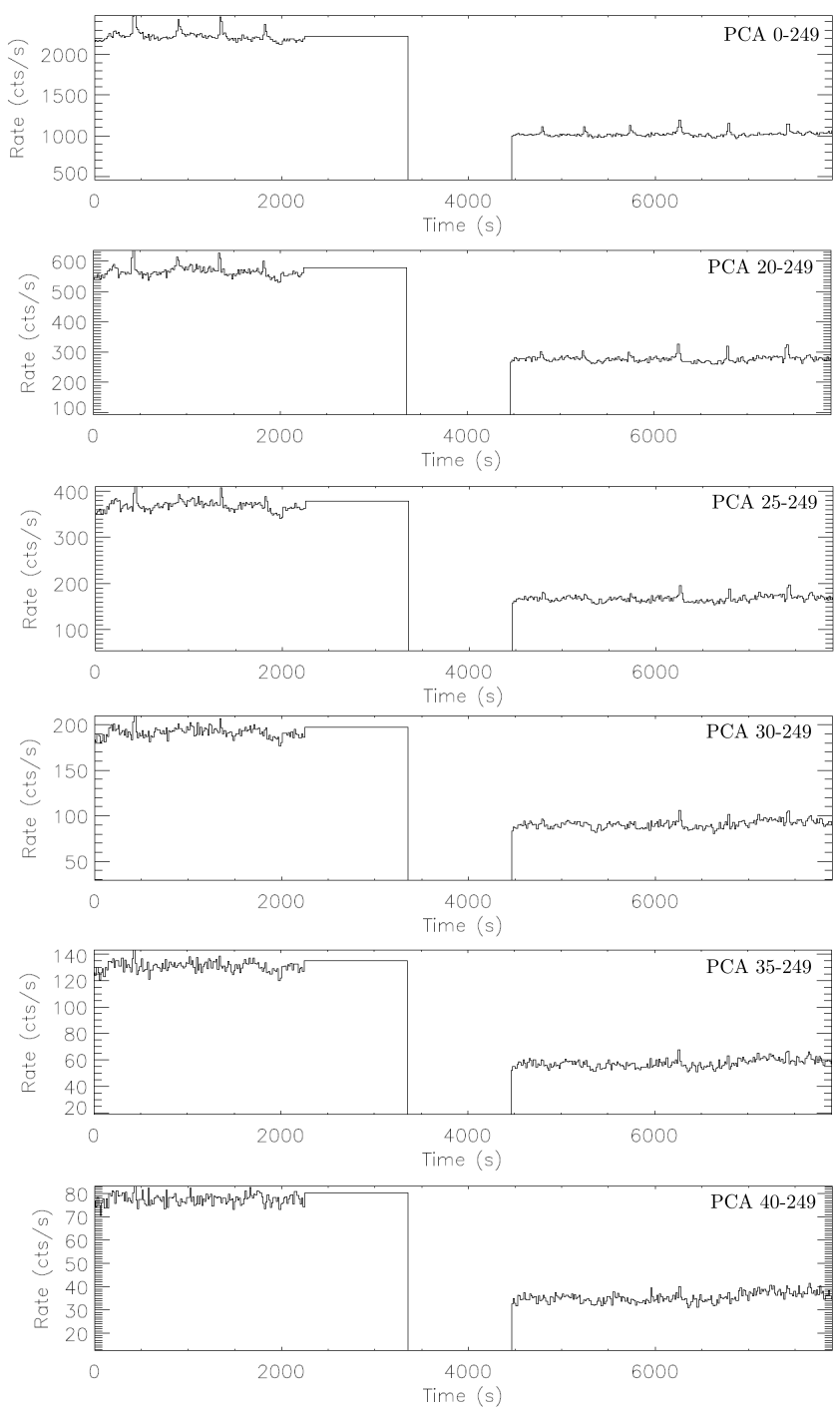}
\caption{The effect of our energy cuts on the light curve of Obs ID 95437-01-10-02. From top to bottom, the cuts are as follows: $0-120$ keV, $8-120$ keV, $10-120$ keV, $12-120$ keV, $15-120$ keV, and $17-120$ keV. Absolute PCA channels used are indicated in the figure.}
\label{fig:lightcurves}
\end{figure*}

\noindent An example of the effect of these cuts on an observation's light curve (observation ID 95437-01-10-02) can be seen in Figure \ref{fig:lightcurves}.

\onecolumn

\begin{landscape} \section{VLF QPO observations}
\label{AppB}

    
	\captionof{table}[Triplets]{Details of the fit parameters of the PDS containing VLF QPOs, found following the \textit{Case 1} analysis (see Section \ref{sec:chap4_case1}). Observation IDs are given in the first column, whereafter the next four columns report the center frequency, full width at half maximum (FWHM), integral power of the Lorentzian, and significance of the Lorentzians used for fitting VLF QPOs. Significance is calculated as the integral of the power of the Lorentzian divided by the negative $1 \, \sigma$ error on this integral. All errors reported in this table are $1 \, \sigma$ errors. The center frequency and FWHM of other Lorentzians used for fitting the remainder of the PDS are reported in the following two columns. In the case that more than one extra Lorentzian were required, the parameters are given on a new line. Obs ID 95437-01-11-06 and 95437-01-12-04 both contain QPOs separate from the VLF QPOs with significances of $4.2 \, \sigma$ and $4.1 \, \sigma$, respectively. These could be second harmonics of the VLF QPOs, and are given on the second line of each of these observations. Obs ID 95437-01-13-04 includes a peak with a significance of $2.4 \, \sigma$, making it a relevant residual at lower frequencies, but excluding it from being a QPO by our specification requirements (see Section \ref{subsec_chap4:selection}). We note that this peak could possibly be the VLF QPO of this observation, but at this point its significance excludes it from being a contender. The fitting of each observation's PDS further includes a power-law component to account for the Poisson noise, in the form $f(x) = Kx^{-\alpha}$; $\alpha$ is set to $0.0$ for each observation, while $K$ is fit - the value of $K$ for each observation can be seen in the last column. Due to the X-ray pulsar nature of the NS in IGR J17480-2446, a very narrow peak corresponding to the $11 \, \textrm{Hz}$ pulsation is visible in some PDS. We fit this peak with a Lorentzian, whereafter we fix the parameters. Obs IDs of observation PDS including this peak is marked with a $\dagger$. Any other parameters that were fixed during PDS fitting is marked with an asterisk.}
	\noindent

    \begin{center}
    \renewcommand{\arraystretch}{1.5}
    \begin{tabularx}{1.2\textwidth}{c|cccc|cc|cc}
    \toprule
    
    & \multicolumn{4}{c}{\textbf{VLF QPO Lorentzian parameters}} & \multicolumn{2}{c}{\textbf{Other Lorentzians}} & & \\
    
    \toprule
    
    \textbf{Obs ID} & \textbf{Freq. (Hz)} & \textbf{FWHM (Hz)} & \textbf{Integral power} & \textbf{Sign.} & \textbf{Freq. (Hz)} & \textbf{FWHM (Hz)} & \textbf{Reduced $\chi^2$} & \textbf{$K$} \\ \toprule
    
    95437-01-02-01$^{\dagger}$  & $0.0101^{+0.0011}_{-0.0008}$ & $0.005 \pm 0.002$ & $\num{7e-4} \pm \num{2e-4}$ & $4.3 \, \sigma$ & $0.003$* & $0.002$* & $1.1$ & $0.036820 \pm \num{7e-6}$\\
    
    95437-01-10-05  & $0.010^{+0.002}_{-0.001}$  & $0.006 \pm 0.003$ & $\num{4.0e-4}^{+\num{1.2e-4}}_{\num{-6e-5}}$ & $6.2 \, \sigma$ & $0.0$* & $\num{7e-8}$* & $0.97$ & $0.020404 \pm \num{5e-6}$ \\
    
    95437-01-11-03  & $0.016 \pm 0.002$  & $0.006^{+0.003}_{-0.002}$ & $\num{4e-4} \pm \num{1e-4}$ & $3.2 \, \sigma$ & $0.0$* & $5e-3$* & $1.0$ & $0.020124 \pm \num{5e-6}$ \\
    
    95437-01-11-06  & $0.020 \pm 0.001$  & $0.005 \pm 0.002$ & $\num{9e-4} \pm \num{2e-4}$ & $3.9 \, \sigma$ & $0.002$* & $0.010$* & $1.0$ & $0.018638 \pm \num{4e-6}$ \\
    
     & & & & & $0.036^{+0.003}_{-0.002}$ & $0.015^{+0.007}_{-0.006}$ & & \\
     
    95437-01-12-04  & $0.0175 \pm \num{6e-4}$ & $0.004^{+0.002}_{-0.001}$ & $\num{1.6e-3} \pm \num{4e-4}$ & $4.2 \, \sigma$ & $0.0$* & $0.006$* & $0.94$ & $0.022009 \pm \num{5e-6}$ \\
    
     & & & & & $0.032^{+0.003}_{-0.002}$ & $0.012 \pm 0.003$ & & \\ 
     
    95437-01-13-04  & $0.029 \pm 0.002$  & $0.009^{+0.004}_{-0.003}$ & $\num{1.2e-3} \pm \num{3e-4}$ & $3.8 \, \sigma$ & $\num{2.4e-3} \pm \num{6e-4}$* & $\num{7.3e-3} \pm \num{8e-4}$* & $0.91$ & $0.038919 \pm \num{9e-6}$ \\
    
    & & & & & $0.0175 \pm \num{8e-4}$ & $0.003^{+0.003}_{-0.002}$ & & \\ 
    
    95437-01-14-00  & $0.014 \pm 0.002$  & $0.007 \pm 0.003$ & $\num{2.2e-3} \pm \num{7e-4}$ & $3.2 \, \sigma$ & $0.0$* & $0.001$* & $0.97$ & $0.13629 \pm \num{3e-5}$ \\
    
    \bottomrule
    
    \end{tabularx}
    \label{table:tab1}
    \renewcommand{\arraystretch}{1.5}
    \end{center}

\end{landscape}

\label{lastpage}
\end{document}